\documentclass{webofc}
\usepackage[varg]{txfonts}   
\woctitle{QUARKS-2016}
\begin{document}
\title{Rare radiative leptonic B-decays}
%
%

\author{\firstname{Anastasiia} \lastname{Kozachuk}\inst{1,2}
          \and
        \firstname{Dmitri} \lastname{Melikhov}\inst{1,3,4}
          \and
        \firstname{Nikolai} \lastname{Nikitin}\inst{1,2,5}
       }

\institute{D.~V.~Skobeltsyn Institute of Nuclear Physics, M.~V.~Lomonosov
Moscow State University, 119991, Moscow, Russia
\and
M.~V.~Lomonosov Moscow State University, Physical Facutly, 119991, Moscow, Russia
\and
Institue for High-Energy Physics, Austrian Academy of Sciences, Nikolsdorfergasse 18, A-1050 Vienna, Austria
\and
Faculty of Physics, University of Vienna, Boltzmanngasse 5, A-1090 Vienna, Austria
\and
A.~I.~Alikhanov Institute for Theoretical and Experimental Physics, 117218 Moscow, Russia
}

\abstract{
We obtain predictions for $B_{(s)}\to e^+e^-\gamma$ and $B_{(s)}\to \mu^+\mu^-\gamma$ decays. All the contributions containing long-distance QCD effects are calculated in the framework of relativistic quark model. The contributions of the light 
vector-meson resonances related to the virtual photon emission from valence quarks 
of the $B$-meson are included. The highest branching ratios for the radiative leptonic B-decays are 
${\cal B}(\bar B^0_s\to e^+e^-\gamma)=18.8\times10^{-9}$ and  
${\cal B}(B^0_s\to \mu^+\mu^-\gamma)=12.2\times10^{-9}$. 
We also give the distribution of the foward-backward asymmetry. 
}
\maketitle
\section{Introduction}
\label{intro}
Rare radiative leptonic $B_{(s)}\to\ell^+\ell^-\gamma$ decays 
are induced by the flavour-changing neutral current transitions $b\to s,d$ which are forbidden at tree level in the Standard Model and are described by penguin and box diagrams, leading to small probabilities of the order of $10^{-8}$ -- $10^{-10}$ (see e.g. \cite{Ali:1996vf}). 
Due to such small values this sector of heavy flavours represented by rare semi-leptonic and radiative decays is expected to be senscitive to contributions of New Physics. Many of the decays have already been studied theoretically and experimentally measured by LHCb, Belle and Babar collaborations, and no considerable descrepancies between experimental results and predictions of the Standard Model have been found. Nevetherless there are several tensions of 
the order of $2-3\sigma$ (see discussion in \cite{Glashow:2014iga,Guadagnoli:2015nra,Guadagnoli:2016erb}): 
The first one is that the ratio ${\cal R}_K={\cal B}(B^+\to K^+\mu^+\mu^-)/{\cal B}(B^+\to K^+e^+e^-)$ is 25\% lower than the SM prediction at 2.6$\sigma$ \cite{Aaij:2014ora, Bobeth:2007dw, Bouchard:2013mia, Hiller:2003js}. In an independent measurement, the braching ratio of $B^+\to K^+\mu^+\mu^-$ is itself 30\% lower that the SM value at 2$\sigma$ \cite{Aaij:2014pli,Aaij:2012vr,Bobeth:2011gi,Bobeth:2011nj,Bobeth:2012vn}. Another inconsistency is probably related to the decay $B_s\to \mu^+\mu^-$. The joint CMS and LHCb measurement of its branching ratio \cite{CMS:2014xfa} also gives the value which is 25\% lower than the SM prediction, but here it is 1$\sigma$ effect only. 
For the branching ratios of $B_s\to \mu^+\mu^-$ and $B_s\to \mu^+\mu^-\gamma$ decays the following relation takes place
\begin{eqnarray}
\frac{{\cal B}(B_s\to \ell^+\ell^-\gamma)}{{\cal B}(B_s\to \ell^+\ell^-)} \,\sim\,\left(\frac{M_{B^0}}{m_{\ell}}\right)^2\frac{\alpha_{em}}{4\pi}, 
\end{eqnarray}
were the squared ratio of masses $(M_{B^0}/m_\ell)^2$ means that the radiative decay $B_s\to \mu^+\mu^-\gamma$ does not have the chirality constraint, $\alpha_{em}$ comes from the photon emission and $4\pi$ in the denominator is the difference between three- and two-particle phase space. For muons one can easily get the estimation $(M_{B^0}/m_\mu)^2\,\sim\,2.5\times 10^3\,\sim\,4\pi/\alpha_{em}$, which means that the branching ratios are approximately equal ${\cal B}(B_s\to \mu^+\mu^-\gamma)\,\sim\,{\cal B}(B_s\to \mu^+\mu^-)$. In fact, ${\cal B}(B_s\to \mu^+\mu^-\gamma)$ is a little bit larger due to additional dynamical effects, such as resonant contributions.

The paper is organized as follows: In Section \ref{sec:2} we discuss the contributions to the decay amplitude $\langle \gamma\ell^+\ell^-|H_{\rm eff}(b\to q \ell^+\ell^-)|B\rangle$. In Section \ref{sec:3} we calculate the transition form-factors via dispersion approach based on constituent quark picture. In Section \ref{sec:4} we give numerical predictions for the branching ratios, differential distributions for the decay rates and forward-backward asymmetry.  


\section{The effective Hamiltonian and the amplitude}
\label{sec:2}
The effective Hamiltonian describing the $b\to q$ ($q=d,s$) weak transition has the form (\cite{Grinstein:1988me, Buras:1994dj})
\begin{equation}
\label{heff}
{\cal H}_{\rm eff}^{b\,\to\, q} = \frac{G_F}{\sqrt{2}} V_{tb} V_{tq}^\ast\, 
\sum_i C_i(\mu) \, O_i(\mu),
\end{equation}
where $G_F$ is the Fermi constant, $C_i$ are the scale-dependent set of Wilson coefficients, and $O_i$ are the basis operators. For $B$ decays the scale parameter $\mu$ is approximately equal to $5$ GeV. The amplitudes of the basis operators between the initial and final states may be parameterized in terms of the Lorentz-invariant form factors. These form factors contain nonperturbative QCD contributions, and therefore their calculation is one the main problems when considering $B\to \ell^+\ell^-\gamma$ decays. 

\subsection{\label{sec:2.1}
Emission of the virtual photon from the penguin}
\begin{figure}[h]
\centering
\includegraphics[width=13cm,clip]{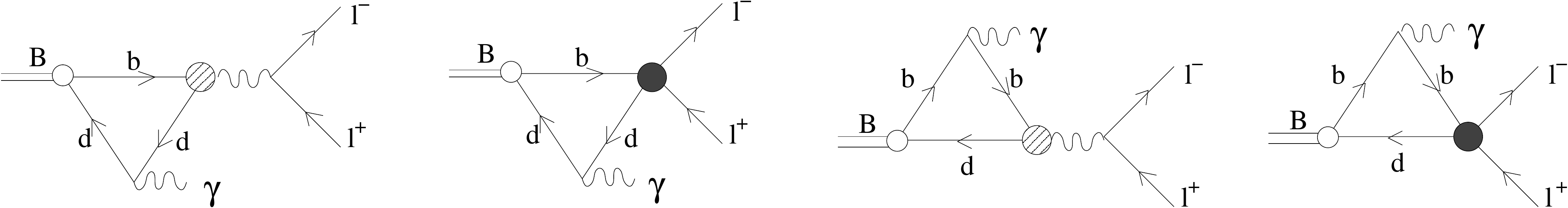}
\caption{\label{fig:1} Diagrams contributing to 
$B\to\ell^+\ell^-\gamma$ discussed in section \protect\ref{sec:2.1}.
Dashed circles denote the $b\to d\gamma$ operator $O_{7\gamma}$.    
Solid circles denote the $b\to d\ell^+\ell^-$ operators $O_{9V}$ and
$O_{10AV}$.}
\end{figure}

The most important part which contains nonperturbative QCD contributions corresponds to the cases, when the real photon is directly emitted from the valence $b$ or $d$ quarks, and the $\ell^+\ell^-$ pair is coupled to the penguin. The effective Hamiltonian in this case takes the form\footnote{Our notations and conventions are: $\gamma^5=i\gamma^0\gamma^1\gamma^2\gamma^3$, $\sigma_{\mu\nu}=\frac{i}{2}[\gamma_{\mu},\gamma_{\nu}]$, $\varepsilon^{0123}=-1$, $\epsilon_{abcd}\equiv\epsilon_{\alpha\beta\mu\nu}a^\alpha b^\beta c^\mu d^\nu$, $e=\sqrt{4\pi\alpha_{\rm em}}$. }  
\begin{eqnarray}
\label{b2qll}
&&H_{\rm eff}^{b\to d\ell^{+}\ell^{-}}\, =\, {\frac{G_{F}}{\sqrt2}}\, {\frac{\alpha_{\rm em}}{2\pi}}\, V_{tb}V^*_{tq}\, \left[\,-2im_b\, {\frac{C_{7\gamma}(\mu)}{q^2}}\cdot\bar d\sigma_{\mu\nu}q^{\nu}\left (1+\gamma_5\right )b\cdot{\bar \ell}\gamma^{\mu}\ell \right.\nonumber\\
&&\left.\qquad\qquad\quad +\, C_{9V}^{\rm eff}(\mu, q^2)\cdot\bar d \gamma_{\mu}\left (1\, -\,\gamma_5 \right)   b \cdot{\bar \ell}\gamma^{\mu}\ell \, +\, C_{10A}(\mu)\cdot\bar d   \gamma_{\mu}\left (1\, -\,\gamma_5 \right) b \cdot{\bar \ell}\gamma^{\mu}\gamma_{5}\ell \right], 
\end{eqnarray} 
and the corresponding diagrams are shown in Fig. \ref{fig:1}. 
The coefficient $C^{\rm eff}_{9V}(\mu, q^2)$ includes long-distance effects related to $\bar cc$ resonances in the $q^2$-channel, where $q^2$ is the invariant mass of the $\ell^+\ell^-$ pair \cite{Kruger:1996dt,Melikhov:1998ws,Melikhov:1997wp}. The $B\to \gamma$ transition form factors of the basis operators in (\ref{b2qll}) are defined according to \cite{Kruger:2002gf}
\begin{eqnarray}
\label{real}
\langle
  \gamma (k,\,\epsilon)|\bar d \gamma_\mu\gamma_5 b|B(p) 
\rangle 
&=& i\, e\,\epsilon^*_{\alpha}\, 
\left ( g_{\mu\alpha} \, pk-p_\alpha k_\mu \right )\,\frac{F_A(q^2)}{M_B}, 
\nonumber
\\
\langle
  \gamma(k,\,\epsilon)|\bar d\gamma_\mu b|B(p)
\rangle
&=& 
e\,\epsilon^*_{\alpha}\,\epsilon_{\mu\alpha\xi\eta} p_{\xi}k_{\eta}\, 
\frac{F_V(q^2)}{M_B},   
\\
\langle
  \gamma(k,\,\epsilon)|\bar d \sigma_{\mu\nu}\gamma_5 b|B(p) 
\rangle\, (p-k)^{\nu}
&=& 
e\,\epsilon^*_{\alpha}\,\left[ g_{\mu\alpha}\,pk- p_{\alpha}k_{\mu}\right ]\, 
F_{TA}(q^2, 0), 
\nonumber
\\
\langle
\gamma(k,\,\epsilon)|\bar d \sigma_{\mu\nu} b|B(p) 
\rangle\, (p-k)^{\nu}
&=& 
i\, e\,\epsilon^*_{\alpha}\epsilon_{\mu\alpha\xi\eta}p_{\xi}k_{\eta}\, 
F_{TV}(q^2, 0).
\nonumber 
\end{eqnarray}
The penguin form factors $F_{TV,TA}(q_1^2,q_2^2)$ are defined as functions of two variables: $q_1$ is the momentum of the photon emitted from the penguin, and $q_2$ is the momentum of the photon emitted from the valence quark of the $B$ meson. We calculate the form factors in the framework of the dispersion approach based on constituent quark picture, the details are presented in Section \ref{sec:3}. 
\subsection{\label{sec:2.2}Emission of the virtual photon from B-meson valence quarks}

\begin{figure}[h]
\centering
\includegraphics[width=7cm,clip]{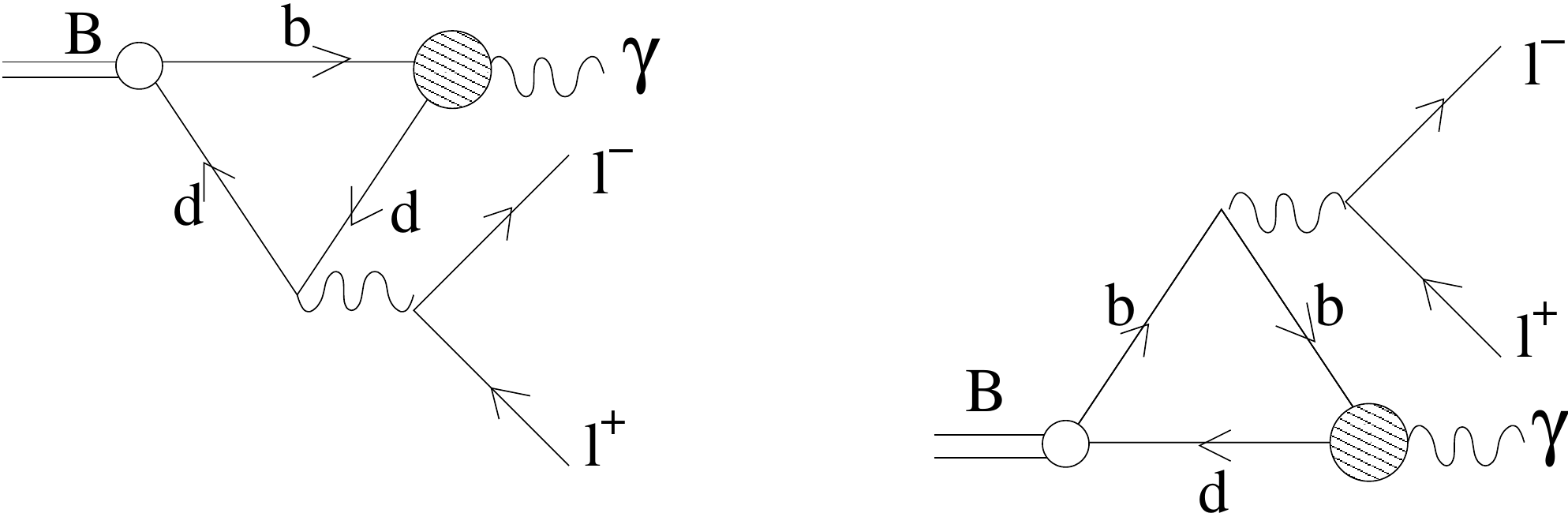}
\caption{\label{fig:2} Diagrams contributing to 
$B\to\ell^+\ell^-\gamma$ discussed in section \protect\ref{sec:2.2}.
Dashed circles denote the $b\to d\gamma$ operator $O_{7\gamma}$.}
\end{figure}

Another process contributing to the amplitude is that with the real photon emitted from the penguin, whereas one of the valence quarks directly emits the virtual photon which then goes into the final $\ell^+\ell^-$ pair. This process is described by the diagrams of Fig. \ref{fig:2}. 
The corresponding amplitude has the same structure as the $C_{7\gamma}$ part of the amplitude in \ref{sec:2.1} with $F_{TA,TV}(q^2,0)$ replaced by $F_{TA,TV}(0,q^2)$. The form factors $F_{TA,TV}(0,q^2)$ at the necessary timelike momentum transfers are not known. The difficulty is connected with appearance of neutral light vector meson resonances, $\rho^0$ and $\omega$ for $B$-decays and $\phi$ for $B_s$-decays, in the physical $B\to \gamma \ell^+\ell^- $ decay region. We calculate the form factors $F_{TA,TV}(0,q^2)$ for $q^2>0$ with the use of gauge-invariant version \cite{Melikhov:2001ew,Melikhov:2003hs} of the vector meson dominance \cite{Sakurai:1960ju,GellMann:1961tg,Gounaris:1968mw}
\begin{eqnarray}
\label{zamena1a}
&&F_{TV,TA}(0, q^2)\, =\, F_{TV,TA}(0, 0)\, -\,\sum_V\,2\,f_V g^{B\to V}_+(0)
\frac{q^2/M_V}{q^2\, -\, M^2_V\, +\, iM_V\Gamma_V},
\end{eqnarray}
where $M_V$ and $\Gamma_V$ are the mass and the width of the vector meson resonance, $g^{B\to V}_+(0)$ are the $B\to V$ transition form factors, defined according to the relations 
\begin{eqnarray}
\langle V(q, \varepsilon)|\bar d\sigma_{\mu\nu} b|
B(p)\rangle
\, =\, i\varepsilon^{*\alpha}\,\epsilon_{\mu\nu\beta\gamma}
\left [ 
g^{B\to V}_+(k^2)g_{\alpha\beta}(p+q)^{\gamma} + 
g^{B\to V}_-(k^2)g_{\alpha\beta}k^{\gamma} + 
g^{B\to V}_0(k^2)p_{\alpha}p^{\beta}q^{\gamma}
\right ] 
\end{eqnarray}
and calculated in \cite{Melikhov:1995xz,Melikhov:1997qk} via relativistic quark model. The leptonic decay constant of a vector meson is given by
\begin{eqnarray}
\langle 0|\bar d \gamma_\mu d|V(\varepsilon, p)\rangle=\varepsilon_\mu M_V f_V.  
\end{eqnarray}

\subsection{\label{sec:2.3}Bremsstrahlung}
\begin{figure}[h]
\centering
\includegraphics[width=3cm,clip]{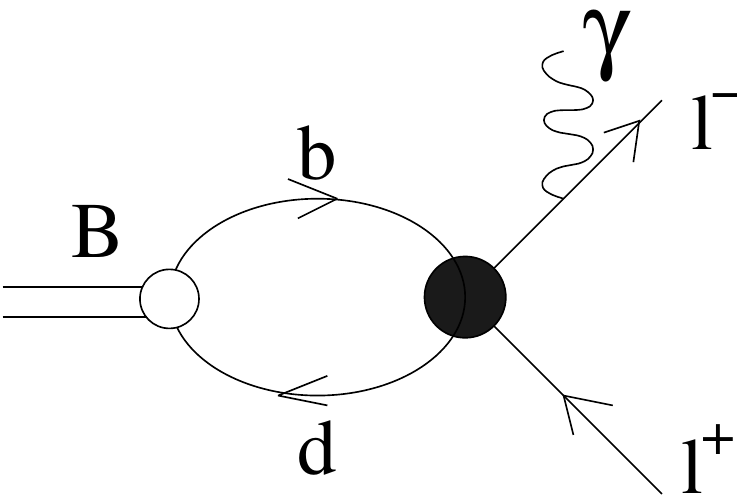}
\caption{\label{fig:3} Diagrams describing photon Bremsstrahlung. Solid
circles denote the operator $O_{10A}$.}
\end{figure}

Fig. \ref{fig:3} gives diagrams describing Bremsstrahlung. The corresponding contribution to the 
$B\to \ell^+\ell^-\gamma$ amplitude  reads 
\begin{eqnarray}
\label{bremsstrahlung}
A_\mu^{\rm Brems}=-i\, e\,\frac{G_F}{\sqrt{2}}\,\frac{\alpha_{\rm em}}{2\pi}\, V^*_{td}V_{tb}\, 
\frac{f_{B_q}}{M_B}\, 2\hat m_{\ell}\, C_{10A}(\mu)\, 
\bar\ell (p_2)
\left [
\frac{(\gamma\epsilon^*)\,(\gamma p)}{\hat t-\hat m^2_{\ell}}\, -\, 
\frac{(\gamma p)\,(\gamma\epsilon^*)}{\hat u-\hat m^2_{\ell}}
\right ]
\gamma_5\,\ell (-p_1),
\end{eqnarray}
$f_B>0$. 
\subsection{\label{sec:2.4}Weak annihilation contribution}

The weak annihilation contribution is given by triangle diagrams of Fig 3. 
\begin{figure}[h]
\centering
\includegraphics[width=5cm,clip]{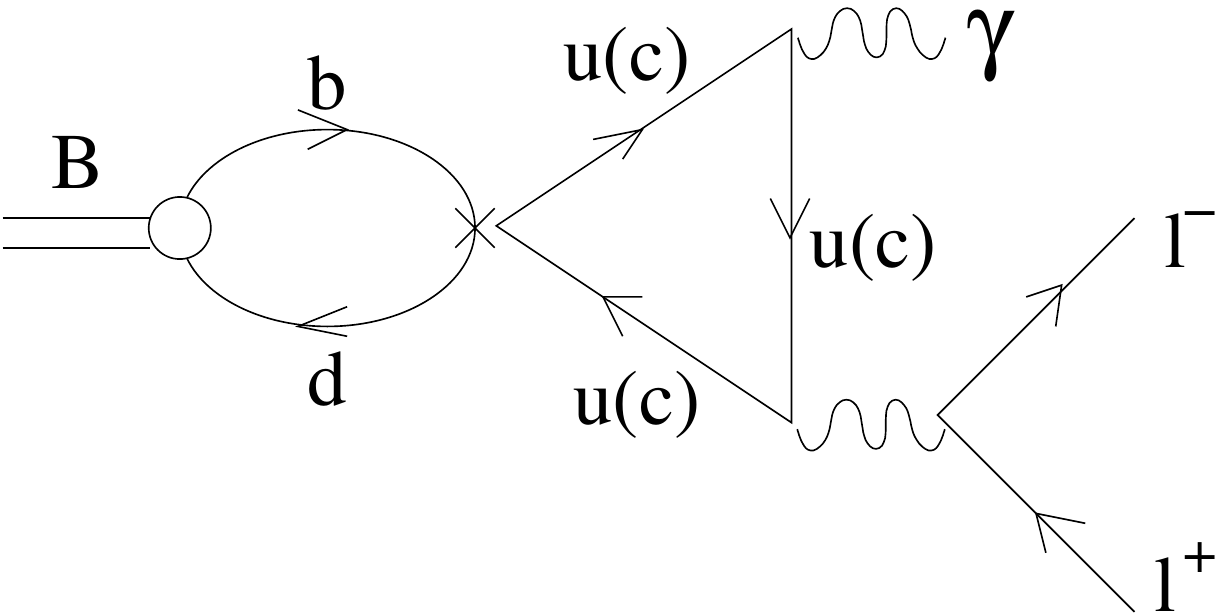}
\caption{\label{fig:4} Weak annihilation diagrams contributing to the process.}   
\end{figure}
One should take into account $u$ and $c$ quarks in the loop. 
The vertex describing the $\bar bd\to \bar QQ$ 
transition ($Q=u,c$) reads
\begin{eqnarray}
\label{heffwa}
H_{\rm eff}^{B\to\bar QQ} = -\,\frac{G_F}{\sqrt{2}}\, a_1\,V_{Qb}V^*_{Qd}
\,\bar d\gamma_{\mu}(1 -\gamma_5)b 
\,\bar Q\gamma_{\mu}(1 -\gamma_5)Q, 
\end{eqnarray}
with $a_1\, =\, C_1\, +\, C_2/N_c$, $N_c$ number of colors \cite{Neubert:1997uc}. 

\section{\label{sec:3} Transition form factors}

We calculate the transition form factors in the framework of relativistic quark model, which is a dispersion approach based on constituent quark picture \cite{Melikhov:1995xz,Melikhov:1997qk}. All hardron observables are given by dispersion representations in terms of hardronic relativistic wave functions and spectral densities of corresponding feynman diagrams with constituent quarks in the loops. For the wave functions we use a Gaussian parametrization $\phi(s)=A(s,\beta)e^{-k^2(s)/(2\beta^2)}$. The simpliest relation can be obtained for a pseudoscalar or vector meson decays constant
\begin{eqnarray} \label{eq:fM}
f_M=\int ds\phi_M(s)\rho(s),
\end{eqnarray}
were $\phi(s)$ is the meson relativistic wave function and $\rho(s)$ is the spectral dencity. The latter is obtained as a direct result of feynman rules for the corresponding feynman diagram. The example of the diagram for a B-meson decay constant is given in Fig. \ref{fig:5}.
\begin{figure}[h]
\centering
\includegraphics[width=3cm,clip]{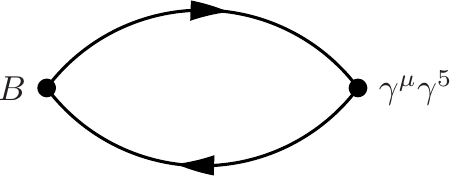}
\caption{\label{fig:5} Feynman diagram corresponding to dispersion representation of B-meson decay constant.}
\label{fig:5}       
\end{figure}
In this work we consider meson-to-photon transitions; the corresponding form factors $F_{V,A,TV,TA}$ may be 
obtained in the form of the spectral represenation 
\begin{eqnarray}
F(q_1,q_2)=\int ds\phi(s)\frac{ds'\Delta(s,s',q_2^2)}{s'-q_1^2},
\end{eqnarray} 
were $q_1$ and $q_2$ are momenta of the emitted photons. The form factors $F_{V,A}$ were calculated 
in \cite{Kozachuk:2015kos}. We now perform the calculation of the form factors $F_{TV,TA}$. 
Each of these form factors contains two contributions corresponding to the diagrams of Fig.~2: 
\begin{eqnarray}
\label{FTV}
F_{TV}\,=\,Q_dF^{(1)}_{TV}(m_d,m_b)+Q_bF^{(1)}_{TV}(m_b,m_d), \\
F_{TA}\,=\,Q_dF^{(1)}_{TA}(m_d,m_b)+Q_bF^{(1)}_{TA}(m_b,m_d)\nonumber.  
\end{eqnarray}
The spectral representations for the form factors in (\ref{FTV}) have the form 
\begin{eqnarray}
F_{TV}^{(1)}(s) \,=\, -\int\limits_{(m_1+m_2)^2}^\infty ds g_2(s, m_1, m_2)-\frac{M_B^2+q^2}{M_B^2-q^2}\int\limits_{(m_1+m_2)^2}^\infty ds g_1(s, m_1, m_2), \\				
F_{TA}^{(1)}(s) \,=\, -\int\limits_{(m_1+m_2)^2}^\infty ds g_2(s, m_1, m_2) - \int\limits_{(m_1+m_2)^2}^\infty ds g_1(s, m_1, m_2),
\end{eqnarray} 
where $m_1$ is the mass of the quark, which emits the photon, $m_2$ is the mass of the spectator, and
\begin{eqnarray}
g_1(s, m_1, m_2) \,=\, \phi_B(s,m_1,m_2) \frac{M_B^2 - q^2}{(s-q^2)^2}\,\Bigg(\,\frac{s + m_1^2 - m_2^2}{2s}\sqrt{\lambda(s, m_1, m_2)} - \\
- m_1^2 \log{\frac{s + m_1^2 - m_2^2 + \sqrt{\lambda(s, m_1, m_2)}}{s + m_1^2 - m_2^2 - \sqrt{\lambda(s, m_1, m_2)}}}\,\Bigg)\,, \nonumber \\
g_2(s, m_1, m_2) \,=\, \phi_B(s,m_1,m_2) \frac{1}{s - q^2}\,\Bigg(\,\sqrt{\lambda(s, m_1, m_2)} - \\
 m_1(m_2-m_1) \log{\frac{s + m_1^2 - m_2^2 + \sqrt{\lambda(s, m_1, m_2)}}{s + m_1^2 - m_2^2 - \sqrt{\lambda(s, m_1, m_2)}}}\,\Bigg)\,\nonumber.
\end{eqnarray}
The model contains only few parameters such as the constituent quark masses and the parameter of the wave function $\beta$. 
These parameters were fixed in \cite{Kozachuk:2015kos} using relations (\ref{eq:fM}) for meson decay constants 
so that our results reproduce the predictions from QCD sum rules and lattice QCD. 

\section{\label{sec:4}Numerical results}
\subsection{Branching ratios}
\begin{figure}[h]
\begin{center}
\begin{tabular}{cc}
\mbox{\centering
\includegraphics[width=6cm,clip]{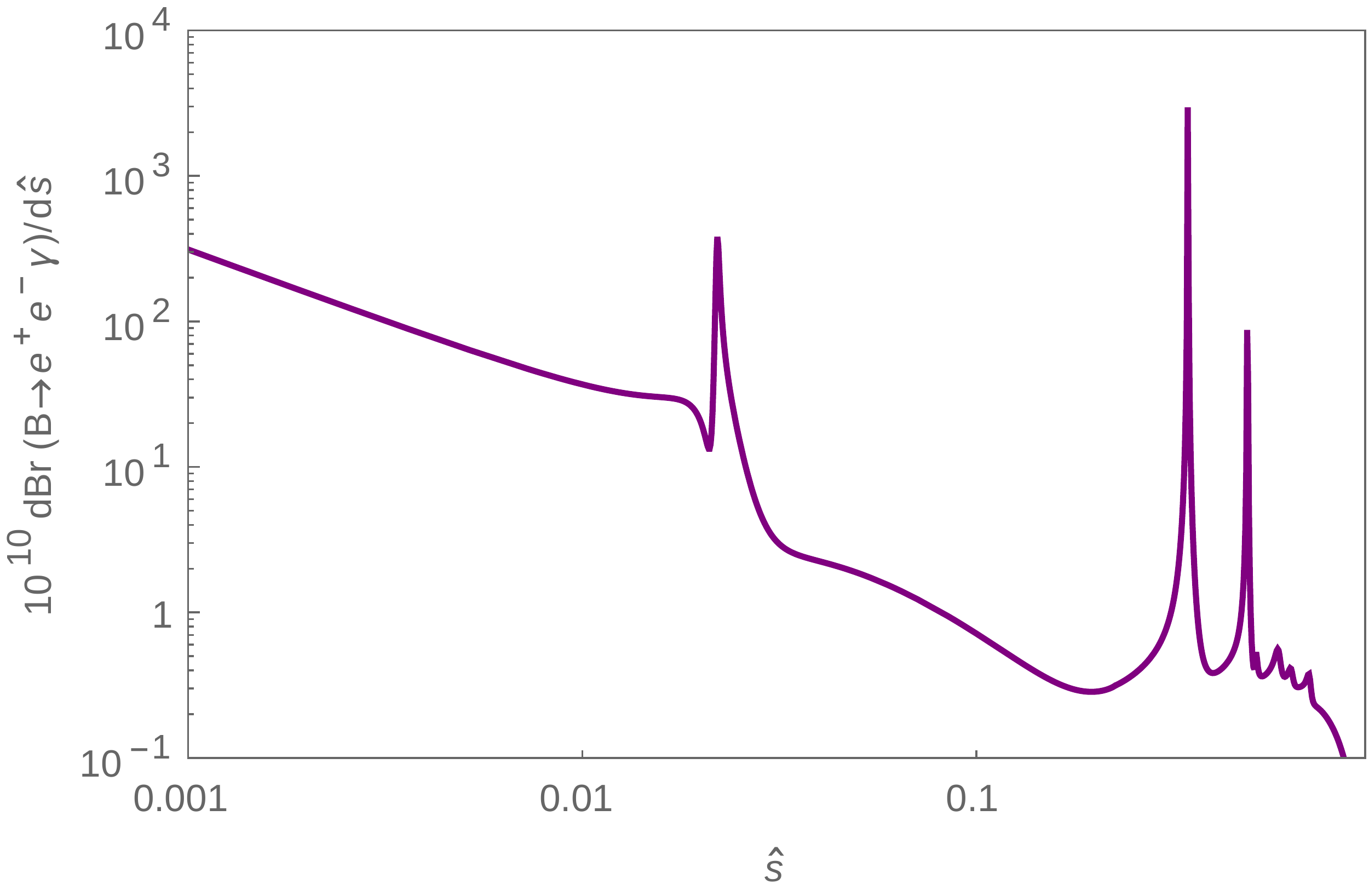}} &
\mbox{\centering
\includegraphics[width=6cm,clip]{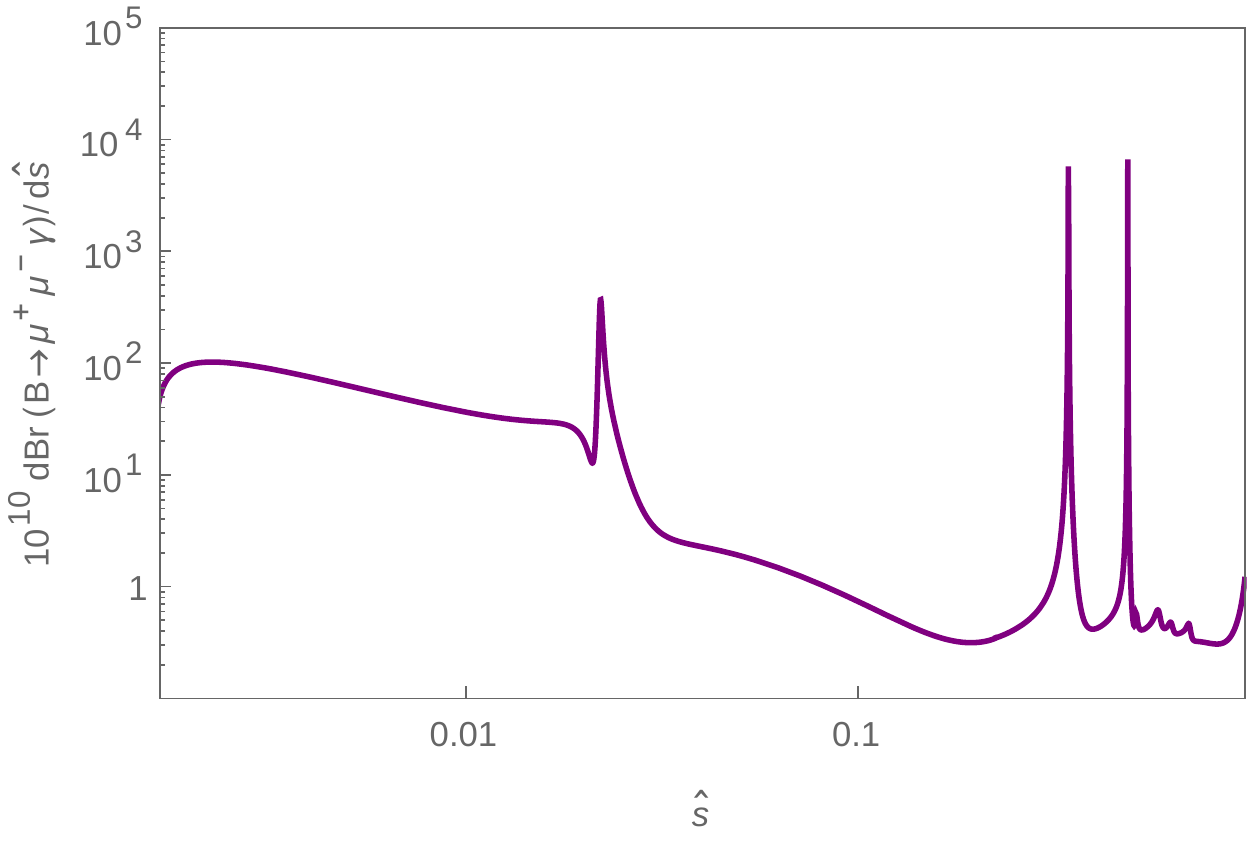}} 
\end{tabular}
\caption{\label{fig:53a} 
Differential branching fractions for $B\to e^+e^-\gamma$ (left) and $B\to \mu^+\mu^-\gamma$ (right) decays.}
\end{center}
\end{figure}

\begin{figure}[h]
\begin{center}
\begin{tabular}{cc}
\mbox{\centering
\includegraphics[width=6cm,clip]{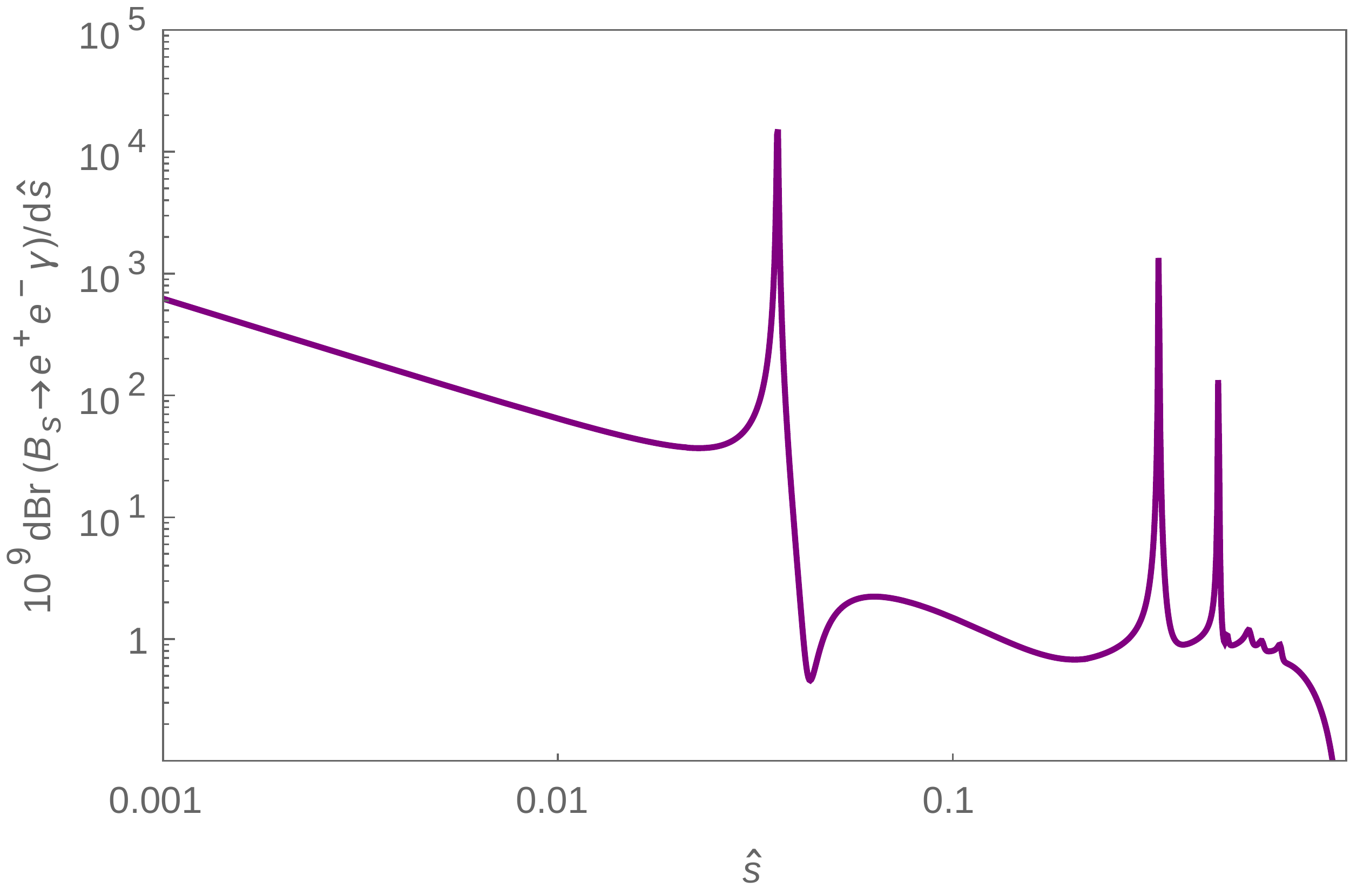}} &
\mbox{\centering
\includegraphics[width=6cm,clip]{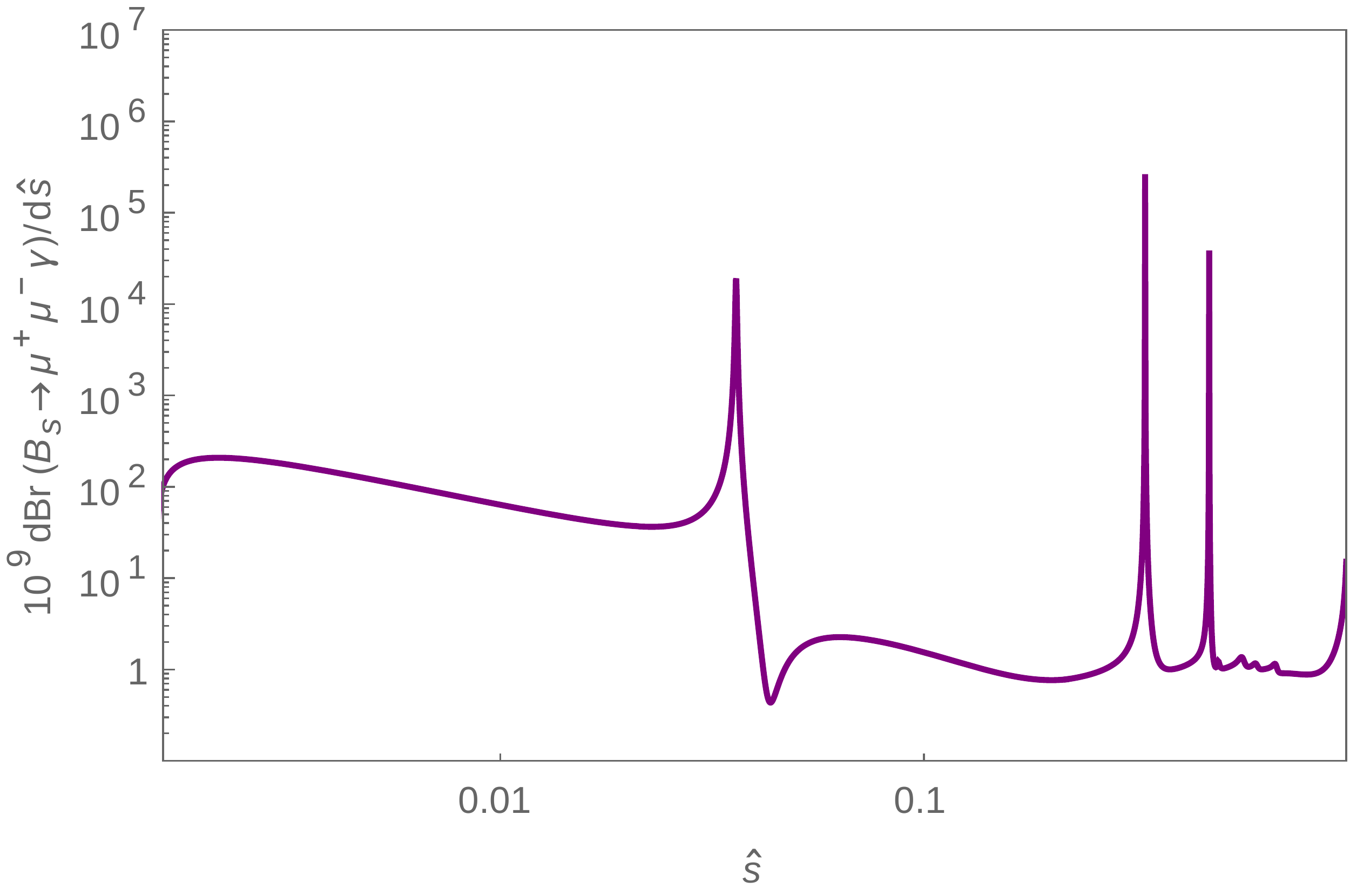}} 
\end{tabular}
\caption{\label{fig:53b} 
Differential branching fractions for $B_s\to e^+e^-\gamma$ (left) and $B_s\to \mu^+\mu^-\gamma$ (right) decays.}
\end{center}
\end{figure}

For numerical estimates we use the following values of Wilson coefficients at $\mu=5$ GeV: \\
$C_1(5\, GeV)=\,0.235$, $C_2(5\, GeV)=\, -1.1$, $C_{7\gamma}(5\, GeV)=\, 0.308$, $C_{10A}(5\, GeV)=4.63$. The $C^{eff}_{9V}(\mu,s)$ evolution including cc-resonances is taken from \cite{Kruger:1996dt,Melikhov:1998ws,Melikhov:1997wp}.
We obtained several distributions for the differential branching fractions, they are shown in Fig.\ref{fig:53a} and \ref{fig:53b}. 
We obtained results for light leptons for the value of the photon energy cut (the minimal photon energy in the $B$-meson rest frame) $E^{\gamma}_{min}=80$ MeV. This particular choice of $E^{\gamma}_{min}$ is connected with the energy resolution of the LHCb detector \cite{LHCB:2000ab,Voss:2009zz}. Our results for the branching ratios are presented in Table \ref{table:res}.

\begin{table}[h]
\centering\begin{tabular}{|c|c|c|c|c|c|}
\hline
                      & this work & \cite{Melikhov:2004mk} & \cite{Wang:2013rfa}  & \cite{Geng:2000fs} & 
                                 \cite{Dincer:2001hu}\\
\hline
\protect\({\cal B}\left ( B\to e^+e^-\gamma\right )\,\times\, 10^{10}\protect\) &
                          4.84 &
                          3.95 &
                          5.8  &
                          1.01 &
                           --\\
\hline
\protect\({\cal B}\left ( B\to \mu^+\mu^-\gamma\right )\,\times\, 10^{10}\protect\) &
                           1.60&
                           1.31&
                            5.8&
                           0.61&
                           --\\
\hline
\protect\({\cal B}\left ( B_s\to e^+e^-\gamma\right )\,\times\, 10^{9}\protect\) &
                           18.8&
                           24.6&
                          16.24&
                           3.29&
                           20\\
\hline
\protect\({\cal B}\left ( B_s\to \mu^+\mu^-\gamma\right )\,\times\, 10^{9}\protect\) &
                           12.2&
                           18.8&
                          16.24&
                           2.00&
                           12\\
\hline
\end{tabular}
\caption{\label{table:res}
Numerical estimates for the branching ratios of $B_{(s)}\to\ell^+\ell^-\gamma$ decays.}
\end{table}

\subsection{Forward-backward asymmetry}

\begin{figure}[h]
\begin{center}
\begin{tabular}{cc}
\mbox{\centering
\includegraphics[width=6cm,clip]{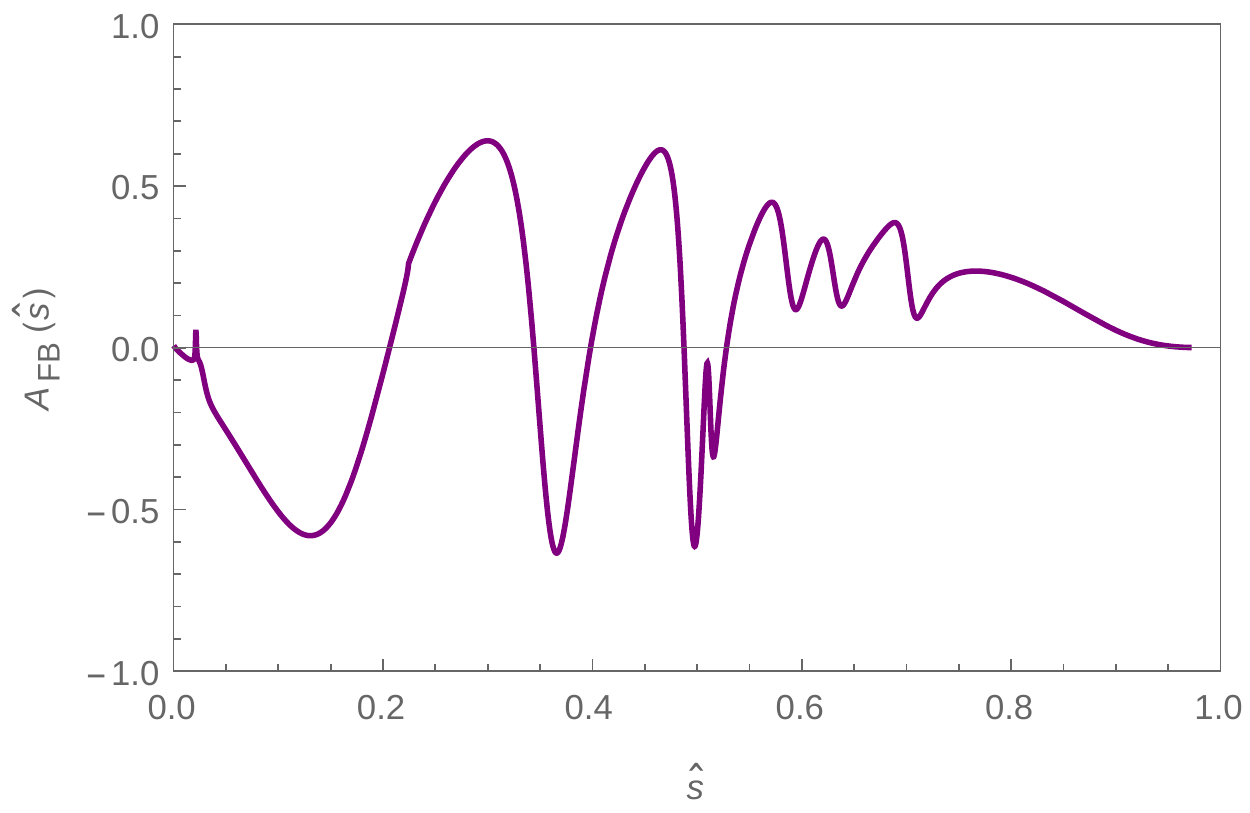}} &
\mbox{\centering
\includegraphics[width=6cm,clip]{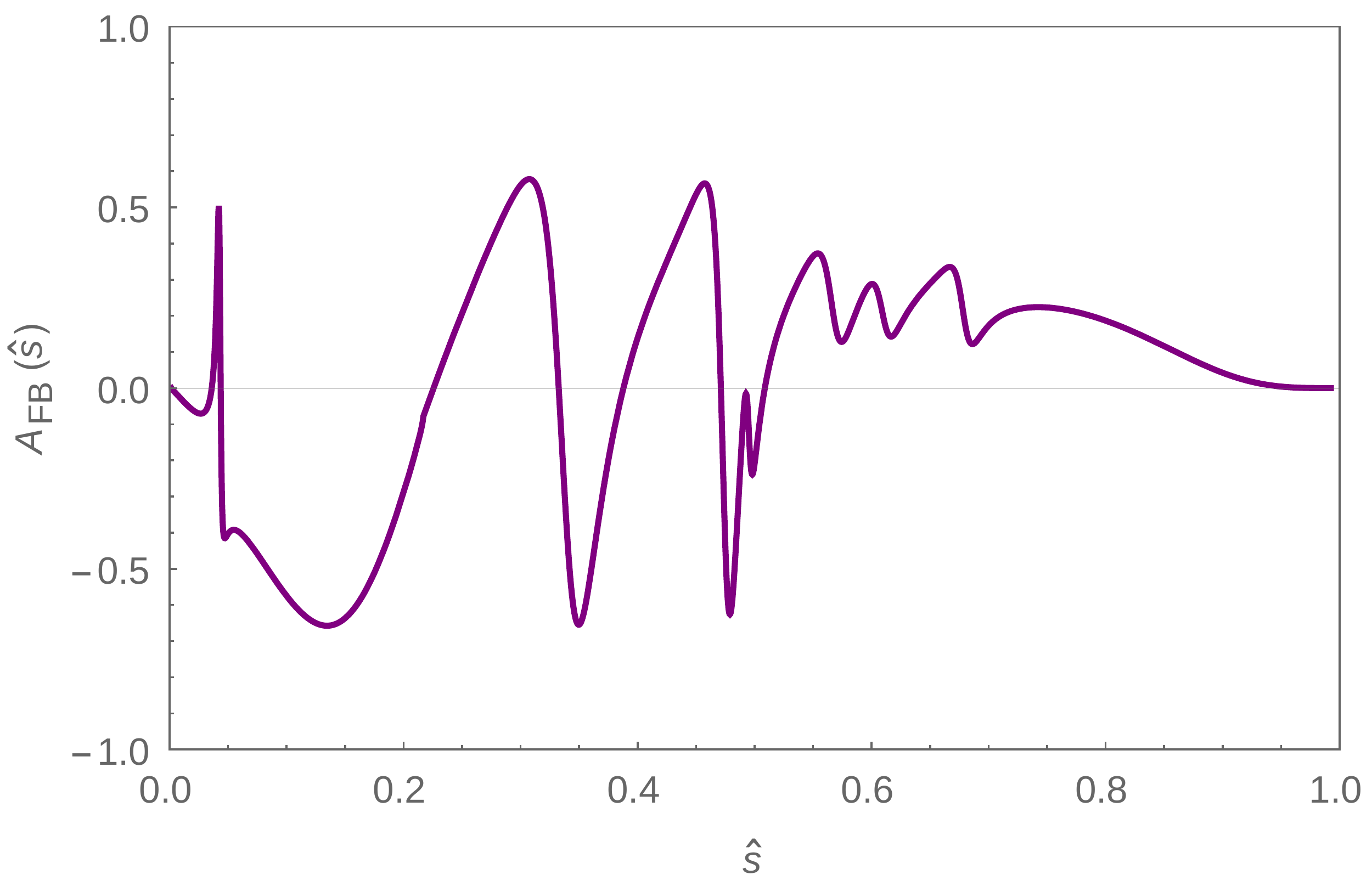}} 
\end{tabular}
\caption{\label{fig:6} 
Forward-backward asymmetry for $B\to \mu^+\mu^-\gamma$ (left) and $B_s\to \mu^+\mu^-\gamma$ (right) decays.}
\end{center}
\end{figure}

We obtained distribution for the forward-backward asymmetry, defined by the relation

\begin{eqnarray}
A_{FB}(\hat{s})\,=\,\frac{\int\limits_0^1d\cos\theta \, \frac{d^2\Gamma(B_{(s)}\to\ell^+\ell^-\gamma)}{d\hat{s} \, d\cos\theta}-\int\limits_{-1}^0d\cos\theta \, \frac{d^2\Gamma(B_{(s)}\to\ell^+\ell^-\gamma)}{d\hat{s} \, d\cos\theta}}{\frac{d\Gamma(B_{(s)}\to\ell^+\ell^-\gamma)}{d\hat{s}}},
\end{eqnarray}

where $\hat{s}\,=\,q^2/M_B^2$,\, $\theta$ is the angle between $\vec{p}$ and $\vec{p_2}$. The distribution is presented in Fig.\ref{fig:6}.

The decay rates and the forward-backward asymmetry were previosly calculated in several works \cite{Geng:2000fs, Dincer:2001hu, Kruger:2002gf, Melikhov:2004mk, Wang:2013rfa, Balakireva:2009kn}. In \cite{Geng:2000fs} and \cite{Dincer:2001hu} not all the contributions were taken into account, and in \cite{Geng:2000fs,Dincer:2001hu, Kruger:2002gf, Melikhov:2004mk} the transition form factors were estimated from symmetry considerations coming from LEET. 
We made direct calculation of the form factors in the framework of the relativistic quark model. 
Our results agree nicely with \cite{Geng:2000fs,Dincer:2001hu}. 
The results \cite{Geng:2000fs,Wang:2013rfa} are based on not fully consistent models for the form factos and therefore do not seem to us convincing. 


\section{Conclusions}
\label{concl}
We obtained predictions for the differential distributions and the branching ratios for the $B_{(s)}\to e^+e^-\gamma$ and $B_{(s)}\to \mu^+\mu^-\gamma$ decays
taking into account the following contributions to the amplitude of the process: the photon emission from the $d(s)$ and the $b$ valence quarks of the $B$-meson, 
the weak annihilation, and the bremsstrahlung. The corresponding form factors were calculated in the framework of the relativistic quark model.

\section{Acknowledgements}
\label{sec:acknownlegements}
The work was supported by grant 16-12-10280 of the Russian Science
Foundation.

\end{document}